%% file: main.tex
\documentclass[footinbib,aps,prl,reprint,superscriptaddress,nopacs]{revtex4-1}

\usepackage{graphicx}
\usepackage{amsmath}
\usepackage{amsfonts}
\usepackage{verbatim}
\usepackage{braket}   
\usepackage{hhline}   
\usepackage[absolute]{textpos}
\usepackage{xcolor}

\input{PhaseFittingPlot.tex}    

\begin{document}
\title{Quantum Phase Estimation with Time-Frequency Qudits in a Single Photon}

\author{Hsuan-Hao Lu}
\thanks{These authors contributed equally to this work.}
\affiliation{School of Electrical and Computer Engineering and Purdue Quantum Science and Engineering Institute, Purdue University, West Lafayette, Indiana 47907, USA}
\author{Zixuan Hu}
\thanks{These authors contributed equally to this work.}
\affiliation{Department of Chemistry, Department of Physics and Purdue Quantum Science and Engineering Institute, Purdue University, West Lafayette, Indiana 47907, USA}
\author{Mohammed S.~Alshaykh}
\affiliation{School of Electrical and Computer Engineering and Purdue Quantum Science and Engineering Institute, Purdue University, West Lafayette, Indiana 47907, USA}
\author{Alexandria J.~Moore}
\affiliation{School of Electrical and Computer Engineering and Purdue Quantum Science and Engineering Institute, Purdue University, West Lafayette, Indiana 47907, USA}
\author{Yuchen Wang}
\affiliation{Department of Chemistry, Department of Physics and Purdue Quantum Science and Engineering Institute, Purdue University, West Lafayette, Indiana 47907, USA}
\author{Poolad Imany}
\affiliation{School of Electrical and Computer Engineering and Purdue Quantum Science and Engineering Institute, Purdue University, West Lafayette, Indiana 47907, USA}
\author{Andrew M.~Weiner}
\email{amw@purdue.edu}
\affiliation{School of Electrical and Computer Engineering and Purdue Quantum Science and Engineering Institute, Purdue University, West Lafayette, Indiana 47907, USA}
\author{Sabre Kais}
\email{kais@purdue.edu}
\affiliation{Department of Chemistry, Department of Physics and Purdue Quantum Science and Engineering Institute, Purdue University, West Lafayette, Indiana 47907, USA}

\date{\today}

\begin{abstract}
The Phase Estimation Algorithm (PEA) is an important quantum algorithm used independently or as a key subroutine in other quantum algorithms. Currently most implementations of the PEA are based on qubits, where the computational units in the quantum circuits are two-dimensional states. Performing quantum computing tasks with higher dimensional states -- qudits -- has been proposed, yet a qudit-based PEA has not been realized. Using qudits can reduce the resources needed for achieving a given precision or success probability. Compared to other quantum computing hardware, photonic systems have the advantage of being resilient to noise, but the probabilistic nature of photon-photon interaction makes it difficult to realize two-photon controlled gates that are necessary components in many quantum algorithms. In this work, we report an experimental realization of a qudit-based PEA on a photonic platform, utilizing the high dimensionality in time and frequency degrees of freedom (DoFs) in a single photon. The controlled-unitary gates can be realized in a deterministic fashion, as the control and target registers are now represented by two DoFs in a single photon. This first implementation of a qudit PEA, on any platform, successfully retrieves any arbitrary phase with one ternary digit of precision.

\end{abstract}

\maketitle

\section{Introduction}

Quantum computation has received enormous attention in recent years with rapid progress in both theoretical and experimental fronts\cite{NAP25196,ladd2010,Preskill2018,Biamonte2017}. The development of quantum algorithms used for quantum computation has been stimulated by the greater availability of more capable quantum devices\cite{Blatt2012, Bernien2017, Dumitrescu2018,Sparrow2018,Kais2014,wang2008,O'Malley2016}. The phase estimation algorithm (PEA) is a key subroutine of several important algorithms such as the Shor’s factorization algorithm\cite{Shor1999} and the Harrow-Hassidim-Lloyd (HHL) algorithm for solving linear systems of equations\cite{Lloyd2009,Cao2012}. PEA has also been developed to find the ground-state energy of a molecular Hamiltonian\cite{Aspuru2005,wang2008,Daskin2011,Aspuru2012}, and experimentally demonstrated on various physical platforms\cite{Lanyon2010,Zhou2013,O'Malley2016,Paesani2017}. Currently most platforms designed for quantum computation are based on quantum bits, or qubits, represented by quantum states in a two-dimensional Hilbert space. The scalability of quantum computation requires representing high-dimensional quantum states with multiple interacting qubits and realizing high-dimensional quantum gates with sequences of one-qubit and two-qubit elementary gates. Due to experimental constraints and environmental noise, both the number of interacting qubits (width) and the length of the gate sequence (depth) limit the capability of quantum computing hardware.

As an alternative to qubits, qudits, represented by quantum states in a $d$-dimensional (with $d$ greater than 2) Hilbert space, has been proposed. Using qudits as the building block can potentially reduce both the width and the depth of quantum circuits, and therefore may offer unique advantages over the conventional qubit systems. Indeed, several benefits of qudits, including higher information coding capacity, stronger non-locality, and enhanced security, have been proposed\cite{Sheridan2010,He2016,Jin2016,Zhang2017,Daboul2003}. Various techniques have demonstrated the required hardware to generate and process qudits by utilizing different degrees of freedom (DoFs) in photons, including orbital angular momentum\cite{Babazadeh2017}, time-bin\cite{Islam2017}, frequency-bin\cite{Kues2017,Lu2018a,Imany2018a}, and hybrid time-frequency bin encoding\cite{Imany2018b,Reimer2019}. Performing quantum simulation and computation with qudits have also been proposed\cite{Lanyon2009,Neeley2009}, but the implementation of a functional quantum algorithm (such as PEA) has not yet been realized on any qudit-based platform. In this work, we experimentally realize a proof-of-principle qudit-based PEA on a photonic platform by encoding two qutrits in a single photon, where the frequency DoF carries one qutrit as the control register, and the time DoF carries another qutrit as the target register. By working with two DoFs in a photon, the controlled-unitary operation required by the PEA is realized within a \emph{single} photon, thus circumventing the undesirable, probabilistic photon-photon interaction\cite{Fiorentino2004,Kagalwala2017,Imany2018b,Reimer2019}. Our qutrit-based implementation is tested on diagonal $3\times 3$ unitary matrices. Eigenphases (i.e. phases associated with the eigenvalues) representable by one ternary digit (given by a single control qutrit) are retrieved with $98\%$ fidelity. For arbitrary eigenphases, we fit their respective photon statistics to theoretical distributions, and minimize the mean squared error. The retrieved phases are all within $7.1\%$ error. In the final section of the paper, we will discuss the possibility of increasing the dimension and complexity of our future system, and show the exploitation of qudits can provide certain advantage in this Noisy Intermediate-Scale Quantum (NISQ) era.

\section{Theory}

\begin{figure}
\includegraphics[width=3.4in]{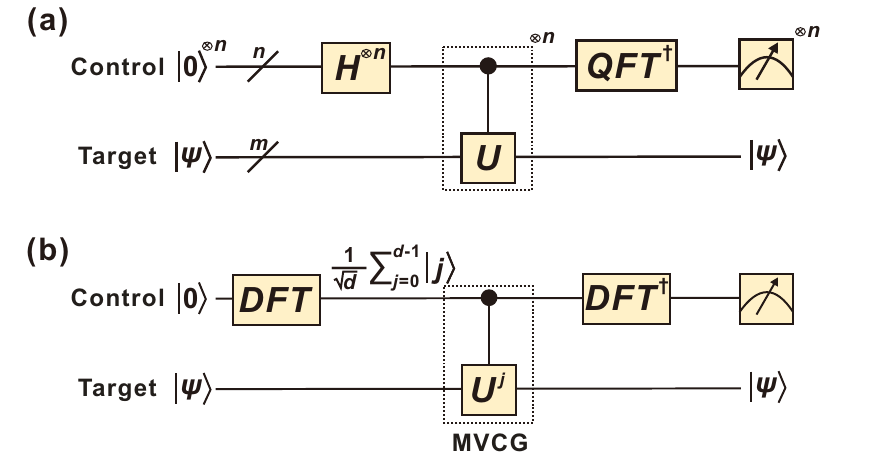}
\caption{(a) Schematic for a general qubit-based PEA using $n$-qubit control and $m$-qubit target states. $H^{\otimes n}$ is a set of Hadamard gates acting on $n$ qubits in parallel. $U$ is an unitary operator operating on the target state. $n$~control-$U$ gates, one for each control rail, are present. $QFT^\dagger$ is the inverse quantum Fourier transform on the $n$-qubit control register. (b) Schematic for a single-qudit-based PEA. $d$-point discrete Fourier transform ($DFT$) is a Hadamard gate generalized to $d$-dimensional state (See Eq.~(\ref{Hadamard})). A multi-value-controlled-gate (MVCG) applies ${{\hat{U}}^{j}}$ to the target when the control is in the $\ket{j}$ state.}
\label{PEA}
\end{figure}

Suppose $\ket{\psi}$  is an eigenstate of a unitary operator $\hat{U}$ with the unknown eigenvalue ${{e}^{i\phi}}$, the phase estimation algorithm can evaluate the phase ($\phi$) with polynomial resource (in terms of the number of qudits and gates needed)\cite{Nielsen2011}. The PEA procedure is illustrated in Figure~\ref{PEA}(a) with a target register to represent $|\psi\rangle $ and a control register to hold the information on $\phi$ which is then extracted by an inverse quantum Fourier transform (QFT)\cite{Nielsen2011}. The dimension of the target register has to match that of $\ket{\psi}$ to fully represent the quantum state, while the dimension of the control register determines the precision of the evaluation of $\phi$. In particular, if we use $n$ $d$-dimensional qudits for the control register then we can evaluate $\phi$ with the precision $2\pi/d^n$. A Hilbert space of dimension $N=2^m=d^n$ can be represented either by $m$ qubits ($d=2$) or $n$ qudits ($d>2$), therefore using a qudit based PEA allows us to achieve the same precision or represent the same $\ket{\psi}$  with fewer number of qudits -- more precisely $m=n\log_2(d)$ implies a $\log_2 (d)$ reduction of the circuit width (number of qudits) required. Using qudits may also reduce the circuit depth by reducing the number of controlled gates used to realize the controlled-$\hat{U}$ operation. The quantum circuit for a qudit based PEA\cite{Cao2011} generalizes the two-value controlled-$\hat{U}$ gate for the qubit case to a multi-value-controlled-gate (MVCG) that applies ${{\hat{U}}^{j}}$ to the target register when the control register is in the $\ket{j}$ state ($j=0,1,...,d-1$). The functionalities of the $n$ two-qubit controlled gates utilized in the conventional qubit-based PEA circuit\cite{Nielsen2011} can be realized with a single MVCG having $2^n$ controlled values, thus reducing the depth of our PEA circuit.


Figure~\ref{PEA}(b) shows the schematic of a qudit-based PEA. The DFT gate here is a $d$-point discrete Fourier transform (DFT) defined as $DFT^{(d)}\ket{j}=\frac{1}{\sqrt{d}}\sum\limits_{k=0}^{d-1} e^{2\pi i(jk/d)}\ket{k}$. The DFT gate can be understood as a qudit generalization of the Hadamard gate to dimensions beyond $d=2$\cite{Wilde2013, Lu2018a}. When operating on a single qudit, both the Hadamard gates and the QFT in Figure~\ref{PEA}(a) are reduced to a single DFT gate. We would like to emphasize that, in this work we use the ``DFT'' to denote a \emph{single}, high-dimensional gate capable of applying the discrete Fourier transform to a single qudit state, while the ``QFT'' denotes the standard quantum algorithm of applying the discrete Fourier transform to a multi-qubit state. Different from the DFT, the QFT often requires a sequence of single-qubit and two-qubit gates to implement. The MVCG then applies $\hat{U}^j$ on the target state conditional on the control state $\ket{j}$ (i.e., $\ket{j}\ket{\psi} \rightarrow \ket{j}\hat{U}^{j}\ket{\psi}$). Finally, the phase kickback mechanism in the PEA\cite{Nielsen2011} allows us to evaluate the $\phi$ by applying an inverse DFT on the control register, and performing measurements in the computational basis. The quantum circuit can determinisitically evaluate the eigenphase $\phi$ for each eigenstate of $\hat{U}$, insofar as $\phi$ can be written exactly with the given precision. If the input state is in superposition of eigenstates instead, performing measurements on the control register will yield probabilistic results, and one can obtain the correct statistics of $\phi$.

As a proof-of-concept implementation, here we limit our dimension to $d=3$ (qutrit) for both the control and target registers, capable of retrieving the eigenphase of a given three-dimensional unitary with ${2\pi}/{3}$ precision. We introduce the three-point DFT gate in its matrix form,

\begin{equation}
    DFT^{(3)}=\frac{1}{\sqrt{3}} \begin{pmatrix} 1 & 1 & 1 \\ 1 & e^{2\pi i/3} & e^{4\pi i/3} \\ 1 & e^{4\pi i/3} & e^{2\pi i/3}
    \end{pmatrix}.
    \label{Hadamard}
\end{equation}
And the unitary ($\hat{U}_1$) of interest in our first demonstration is simply a Pauli-Z gate generalized to the qutrit space:

\begin{equation}
    \hat{U}_1= \begin{pmatrix} 1 & 0 & 0 \\ 0 & e^{2\pi i/3} & 0 \\ 0 & 0 & e^{4\pi i/3}
    \end{pmatrix}
\label{unitary}
\end{equation}
where the eigenphases $0$, ${2\pi}/{3}$, and ${4\pi}/{3}$ can be exactly represented with a single ternary digit expansion.

\section{Experimental results}

\begin{figure}
\includegraphics[width=3.4in]{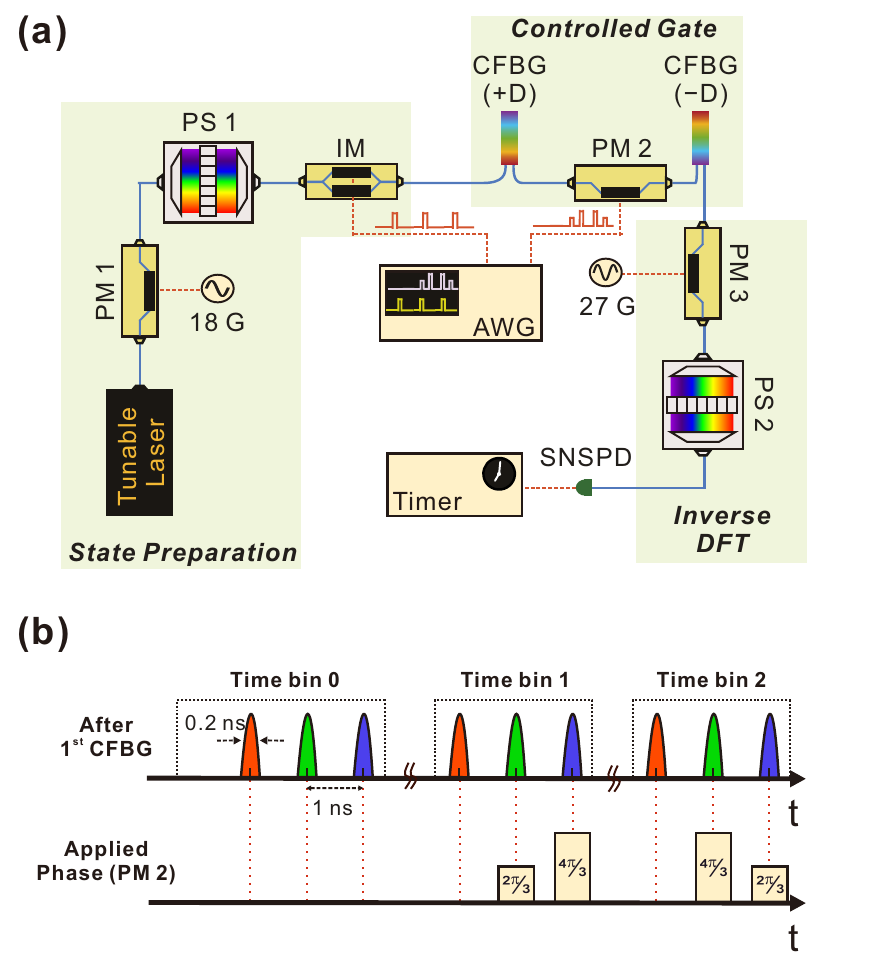}
\caption{(a) Experimental setup. (b) Implementation of controlled-phase gate. See text for details. (PM/IM: Electro-optic phase/intensity modulator; PS: Fourier-transform pulse shaper; CFBG: Chirped fiber bragg grating; SNSPD: Superconducting nanowire single-photon detector; AWG: Arbitrary waveform generator. Both radio-frequency oscillators (18 and 27~GHz) are synchronized to the 10~MHz reference clock of the AWG.) }
\label{Setup}
\end{figure}

In this experiment, we leverage the well-established techniques and fiber-optic components developed for optical communication and wavelength division multiplexing to create and manipulate high dimensional quantum states for PEA implementation. Figure~\ref{Setup}(a) provides a schematic of the setup, which can be decomposed into three stages: state preparation, high-dimensional controlled operation, and measurement on the control qudit. 

To prepare an equi-amplitude superposition of frequency qutrit as the control register, we send a continuous-wave (CW) laser source operating in the C-band through a phase modulator (PM1) driven at 18~GHz, which creates a total number of ${\sim}10$ frequency bins with a spacing of 18~GHz. Subsequently, a pulse shaper (PS1) is programmed to filter out all but three equi-amplitude frequency bins, now with a frequency spacing ($\Delta f$) of 54~GHz. Note that since the controlled gate in our proposed setup is a one-photon operation, the input photon number statistics have no impact on the operation, thus coherent states can be used instead of true single photons as the input.

To prepare the target qutrit state, we employ an intensity modulator (IM) driven by an arbitrary waveform generator (AWG), and carve out three narrow time bins each with a 6~ns spacing, a 24~ns repetition period, and a full width at half maximum of $\sim$0.2~ns which broadens the frequency-bin line-width to 2.2 GHz. As our unitary matrix of interest (Eq \ref{unitary}) is diagonal, the target qutrit eigenstates are single time bins. Thus, we choose to treat each time bin as an independent eigenstate. Each experimental trial can be thought of as three separate measurements made in quick (6~ns) succession. Considering only one of the time bins (eigenstates) at a time, the state after the state preparation stage can be written as:
\begin{equation}
\ket{\psi}_{in} \propto \sum_{j=0}^{2} \ket{j}_f \otimes \ket{\tau}_t
\end{equation}
where $\tau =\{0, 1, 2\}$ denotes which time bin is chosen to operate. The controlled gate itself consists of a phase modulator (PM2) sandwiched between two chirped fiber Bragg grating (CFBG). The first CFBG has a dispersion of 2~ns/nm imparting a frequency-dependent delay which splits each time bin into 3 daughter time bins, each of which corresponds to one frequency mode [red, green and blue pulses in Figure~\ref{Setup}(b)]. The spacing between daughter time bins ($\Delta t$) is 0.9~ns, which is larger than the time-bin coherence time ($\sim$0.2 ns) and its product with the frequency-bin spacing (54 GHz) exceeds the Fourier transform limit (i.e., $\Delta f\Delta t> 1$), allowing independent manipulation of the time and frequency DoFs. Using the AWG, we program the phase modulator to apply the unitary $U^j$ defined in Eq.~(\ref{unitary}) to time-bin states conditional on the frequency-bin state $\ket{j}_f$. The second CFBG with an opposite dispersion of $-2$~ns/nm cancels the first dispersion module and recombines the three daughter time bins back into a single indistinguishable time bin. After the application of MVCG, we can obtain an output state 
\begin{equation}
\ket{\psi}_{out} \propto \sum_{j=0}^{2} e^{\frac{i2\pi j \tau}{3}}\ket{j}_f\otimes\ket{\tau}_t
\label{kickback}
\end{equation}
Note that the phases applied to the time-bin state are now attached to the control register, a process called ``phase kickback."


An ideal three-point inverse DFT gate performs the following transformation: 
\begin{equation} 
    \frac{1}{\sqrt{3}} \sum_{j=0}^{2} e^{\frac{i2\pi j \tau}{3}}\ket{j}_f \xrightarrow{DFT^{-1}} \ket{\tau}_f
\label{trit_IQFT2}
\end{equation}
and thus applying inverse DFT on the control state allows us to read out the phase based on the detection pattern in the logical basis. Detection in output state $\ket{\tau}_f$ indicates the retrieved phase $\tilde{\phi}$ equals to $2\pi\tau/3$ ($2\pi\times0.\tau_{3}$ in ternary expression). Recently a near-deterministic, three-dimensional DFT for frequency-encoded qutrit has been demonstrated with near-unity fidelity, utilizing a quantum frequency processor circuit\cite{Lu2018b} consisting of two electro-optic phase modulators and one pulse shaper. Due to equipment availability, we elect to implement a simpler, probabilistic~\footnote{A single phase modulator will necessarily scatter input photons out of the computational space, thus make the gate probabilistic. See \cite{Lu2018a,ImanyHOM,Imany2018a} for detailed discussions.} version of inverse DFT using a single phase modulator (PM3), capable of performing the equivalent functions in a multi-shot fashion. The control state, consisting of three frequency bins with 54~GHz spacing, is phase modulated by a 27~GHz sine waves to create frequency sidebands. We fine-tune the modulation index to 1.843~rad such that each frequency bin projects onto the central bin $\ket{1}_f$ with equal probability. We utilize another pulse shaper (PS2) as a bandpass filter to pick out this overlapped bin, and then route to a superconducting nanowire single-photon detector (SNSPD) for measurement. Since the output now consists of projections from all three frequency bins, the measured counts will reflect the relative phases due to interference. 

Given a control register in the state $\propto \sum_{j=0}^{2} \ket{j}_f$ (LHS of Eq.~(\ref{trit_IQFT2}) when $\tau=0$) as the input of the PM3, after frequency mixing we have maximum photon counts in the overlapped bin due to constructive interference. The other two orthogonal states will instead experience destructive interference and thus contribute no photon counts. This operation is equivalent to the transformation described in Eq.~(\ref{trit_IQFT2}) for $\tau=0$, namely the projection onto $\ket{0}_f$. We can also tune the delays between the input photons and the electrical drive on PM3, such that the other two transformations (Eq.~(\ref{trit_IQFT2}) for $\tau=1, 2$) can be achieved. Different delay settings can be achieved by introducing an additional pulse shaper prior to frequency mixing, or a radio-frequency phase shifter to impart the required delay. Here we choose to lump this function into PS1 in the state preparation stage to reduce the insertion loss and the complexity of the system. To avoid any confusion, for the rest of the paper we name the three delay settings required to realize the equivalent inverse DFT functions simply as ``projection onto $\ket{0}_f$, $\ket{1}_f$ and $\ket{2}_f$", respectively.



\begin{figure}
\includegraphics[width=3.4in]{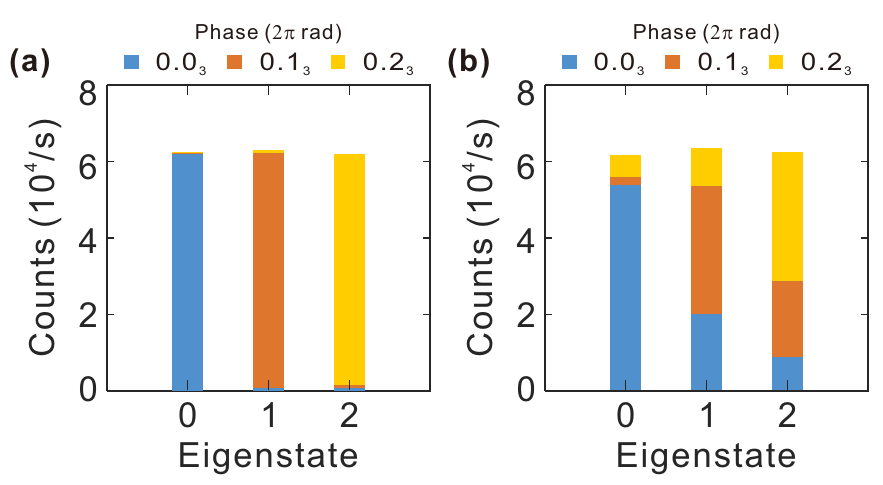}
\caption{Experimental results for implementing (a) $\hat{U}_1$ (Eq.~\ref{unitary}), and (b) $\hat{U_2}$ (Eq~\ref{Unitary2}). Color of the bars corresponds to number of photon counts registered after inverse DFT projection onto  $\ket{0}_f$ (blue), $\ket{1}_f$ (red), and $\ket{2}_f$ (yellow). Photon counts are recorded over 1 second.
}
\label{Results}
\end{figure}

Under each delay setting, we measure the photon counts in three time bins recorded over 1 second. We note that these time bins are widely spaced and do not interfere, hence, this measurement can be considered as three independent measurements of each eigenstate in series. As shown in Eq.~\ref{kickback}, the phase attached to the control register after the MVCG matches the inverse DFT transformation described in Eq.~\ref{trit_IQFT2}, thus we have 
\begin{equation} 
    \sum_{j=0}^{2} e^{\frac{i2\pi j \tau}{3}}\ket{j}_f \otimes \ket{\tau}_t \xrightarrow{DFT^{-1}\otimes I} \ket{\tau}_f \otimes \ket{\tau}_t
\end{equation}
which shows for time-bin $\ket{\tau}_t$ as the input target, ideally we will only obtain photon counts after projecting the control register onto $\ket{\tau}_f$. Figure~\ref{Results}(a) shows the experimental results for estimating the eigenphase of $\hat{U}_1$. For each target eigenstate, we stack three color-coded vertical bars in a single slot to represent the registered counts for different frequency projections. The total number of counts remain stable across three successive measurements, and most of the counts for eigenstate $\ket{\tau}_t$ are recorded after projection onto $\ket{\tau}_f$. The results match our prediction, as all three eigenphases for $\hat{U}_1$ can be represented with exactly one ternary digit, and thus the phase can be retrieved deterministically. The fidelity of this measurement, here defined as the ratio of photon counts registered at the correct output to the total number of received photons, is $98 \pm 1\%$.

For the second part of the experiment, we reprogram our MVCG operation by applying a different temporal phase mask imparted by the AWG to implement another unitary $\hat{U_2}$: 
\begin{equation}
    \hat{U_2}=\left( \begin{matrix}
   1 & 0 & 0  \\
   0 & {{e}^{i{0.351\pi }\;}} & 0  \\
   0 & 0 & {{e}^{i{1.045\pi }\;}}  \\
\end{matrix} \right).
\label{Unitary2}
\end{equation}
Note that two of the eigenphases are no longer integer multiples of $2\pi/3$, or namely, the phase attached onto the control frequency bins (after phase kickback mechanism) does not match the inverse DFT to transform into a single logical state. Figure~\ref{Results}(b) shows the measured counts for all eigenstates under different inverse DFT settings. For the first eigenstate, its corresponding phase is $0$ and hence most of the counts are still registered in $\ket{0}_f$. The other two eigenstates, as discussed above, possess phases which cannot be accurately retrieved given a single ternary digit precision, and thus the counts are distributed over different projections. Following the conventional PEA approach, we report the retrieved phase ($\tilde{\phi}$) based on the projection with the highest number of counts. For eigenphases $\phi$ equal to $0.351\pi$ and $1.045\pi$, the corresponding $\tilde{\phi}$ are $2\pi/3$ and $4\pi/3$, respectively. In the following section, we will discuss whether more information can be extracted from the counts distribution shown in Figure~\ref{Results}, given (i) the input state is already prepared in the eigenstate, and (ii) an ample amount of counts are registered for further analysis before the system decoheres.

\section{Discussion}
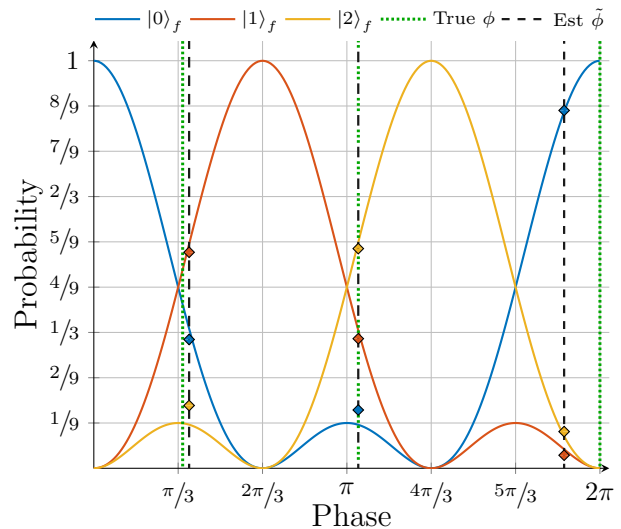
\begin{figure}
    \centering
    \begin{tikzpicture}[x=1cm,y=6cm, baseline=(current bounding box.center), thick,scale=1, every node/.style={scale=1.3}]
        \draw (current bounding box.north east) -- (current bounding box.north west) -- (current bounding box.south west) -- (current bounding box.south east) -- cycle;
    \end{tikzpicture}
    \PhaseFitting
    \caption{The three curves represent the probability of the control qutrit collapsing to bin $\ket{n}_f$ for $n=\{0$(blue)$, 1$(red)$, 2$(yellow)$\}$ for any eigenphase corresponding to the eigenstate input of a PEA circuit. The normalized experimental photon counts for all three eigenstates of $\hat{U}_2$ are fitted to -- and plotted at -- the estimated phase ($\tilde{\phi}$) which minimizes mean squared error (Eq~\ref{Eq:min}). $\tilde{\phi}$ is denoted by the position of the markers and the black dashed line; the location of the true phase ($\phi$) for each eigenstate is noted with a green dashed line. The curves are described by Eq \ref{Eq:controlProb}; the plotted values can be found in Table~\ref{tab:Results}.}
    \label{fig:PhaseFitting}
\end{figure}

When the input target register of a PEA circuit is an eigenstate with a corresponding eigenphase~$\phi$, the probability for the qutrit output control state to collapse to $\ket{n}$, where $n = \{0,1,2\}$, is
\begin{align}
    C(n,\phi)=\frac{1}{9}\left|1+e^{i(\phi -\frac{n2\pi}{3})}+e^{i2(\phi -\frac{n2\pi}{3})}\right|^2.
\label{Eq:controlProb}
\end{align}
All three $C(n,\phi)$, for $n = \{0,1,2\}$, are plotted in Figure~\ref{fig:PhaseFitting}. Observe that for each $\phi$, the ordered set $\{C(0,\phi),C(1,\phi),C(2,\phi)\}$ is unique. Now let $E_0$, $E_1$, and $E_2$ be the measured, normalized ($\sum E_n = 1$) photon counts projected, respectively, onto $\ket{0}_f$, $\ket{1}_f$, and $\ket{2}_f$. The phase we estimate from our measurement, denoted $\tilde{\phi}$, is the phase which minimizes the mean squared error between the measured and theoretical probabilities:
\begin{equation} \label{Eq:min}
    \min_{\tilde{\phi}} \sum_{n=0}^2 (E_n - C(n,\tilde{\phi}))^2
\end{equation}
The estimated phases for $\hat{U}_1$ (Eq.~\ref{unitary}) and $\hat{U}_2$ (Eq.~\ref{Unitary2}) are shown in Table \ref{tab:Results}. The results for $\hat{U}_2$ are plotted in Figure~\ref{fig:PhaseFitting} alongside the three $C(n,\phi)$ curves of Eq.~\ref{Eq:controlProb}. The largest error in $\tilde{\phi}$ is $7.1\%$, and the error is less than $3\%$ in all other cases. Our photonic system's ability to execute large number of trials enables this statistical approach to phase estimation. Agreement between estimated and true phase can be used to quantify error in the experimental setup; however, because the statistical approach requires an eigenstate input, it should not be viewed as a standalone method for determining an unknown phase. To obtain a more precise phase estimate, where the input need not be an eigenstate and the phase is not restricted to a value representable by a single ternary digit, an iterative PEA, explained below, must be implemented.

\renewcommand{\arraystretch}{1.45}
\begin{table}[]
    \centering
    \begin{tabular}{|c||c|c|c|}
    \hline
    \multicolumn{4}{|c|}{$\hat{U}_1$}   \\ \hline
    \textbf{Eigenstate}             
    & $\boldsymbol{\ket{0}_t}$        & $\boldsymbol{\ket{1}_t}$          & $\boldsymbol{\ket{2}_t}$       \\ \hline \hline 
    $\boldsymbol{E_0}$                           
    & $.9948\pm.0004$    & $.0101\pm.0004$      & $.0122\pm.0005$     \\ \hline
    $\boldsymbol{E_1}$                           
    & $.0023\pm.0002$    & $.9805\pm.0009$      & $.0120\pm.0005$     \\ \hline
    $\boldsymbol{E_2}$                           
    & $.0029\pm .0002$   & $.0094\pm.0004$      & $.9758\pm.0010$     \\ \hline
    \textbf{True Phase}, $\phi$     
    & $0$                & $2\pi/3$           & $4\pi/3$        \\ \hline
    \textbf{Est. Phase}, $\tilde{\phi}$    
    & $1.972\pi$        & $.612\pi$           & $1.394\pi$       \\ \hline
    \textbf{Error}, $\frac{|\phi - \tilde{\phi}|}{2\pi}$  
    & $1.4\%$           & $2.7\%$             & $3.0\%$             \\ \hline
    \hline \hline
    \multicolumn{4}{|c|}{$\hat{U}_2$}   \\ \hline
    \textbf{Eigenstate}             
    & $\boldsymbol{\ket{0}_t}$        & $\boldsymbol{\ket{1}_t}$          & $\boldsymbol{\ket{2}_t}$       \\ \hline \hline
    $\boldsymbol{E_0}$                           
    & $.878\pm.002$      & $.316\pm.003$        & $.143\pm.002$     \\ \hline
    $\boldsymbol{E_1}$                           
    & $.032\pm.001$      & $.530\pm.003$        & $.318\pm.003$     \\ \hline
    $\boldsymbol{E_2}$                           
    & $.090\pm .002$     & $.154\pm.002$        & $.539\pm.003$     \\ \hline
    \textbf{True Phase}, $\phi$     
    & $0$                & $.3511\pi$           & $1.045\pi$        \\ \hline
    \textbf{Est. Phase}, $\tilde{\phi}$   
    & $1.859\pi$        & $.377\pi$           & $1.045\pi$       \\ \hline
    \textbf{Error}, $\frac{|\phi - \tilde{\phi}|}{2\pi}$  
    & $7.1\%$           & $1.3\%$             & $0.0\%$             \\ \hline
    \end{tabular}
    \caption{Normalized photon counts and comparison of true phase $\phi$ and experimentally estimated phase $\phi'$ for each eigenstate of $\hat{U}_1$ (Eq.~\ref{unitary}) and $\hat{U}_2$ (Eq.~\ref{Unitary2}). Photon counts normalized from results in Figure~\ref{Results}.}
    \label{tab:Results}
\end{table}

The next steps for our qudit-based PEA are (i) implementing arbitrary unitaries (i.e. non-diagonal) in addition to increasing the qudit dimension ($d>3$); and (ii) increasing the digits of precision for evaluating the phase. For Step (i) we choose to work with frequency and time DoF in photons, since we can take advantage of their inherent high dimensionality to encode more quantum information in a single qudit. For example, our group has recently demonstrated a two-photon four-party GHZ state by encoding two 32~dimensional qudits in each photon.\cite{Imany2018b} In addition, the recipe of constructing high-dimensional quantum gate, though still relatively limited, has been proposed and experimentally realized on both time-bin and frequency-bin platforms. Scalable processing of time-bin qudits has been proposed using a cascade of electro-optic phase modualtor and coded fiber Bragg grating pairs.\cite{Lukens2018} Quantum state tomography of time-bin quqarts ($d=4$) has also been realized with cascaded Mach-Zehnder interferometers fabricated on planar light-wave circuit.\cite{Ikuta2017} And finally, our group has been involved in the design and construction of a quantum frequency processor\cite{Lu2018b} consisting of a series of phase modulators and pulse shapers, found capable of implementing qudit transformations with favorable component requirements. Though eventually to implement a general MVCG operation still demands careful design and perhaps exploitation of other DoFs to realize the controlled operation, the basic formula is ready to be explored.

For Step (ii), our current setup uses a single control qudit to estimate the phase with $2\pi/d$ precision (in this experiment, $d=3$). To achieve higher precision phase estimation we can either increase the number of control qudits or implement an iterative PEA.\cite{Dobsicek2007} The iterative PEA is the more viable approach for photonic systems, as it avoids the difficulty in manipulating and interacting multiple photon qudits. Using only one $d$-dimensional qudit as the control register, the iterative PEA can evaluate the phase with $2\pi/d^n$ precision by running $n$ iterations of a modified single-qudit PEA algorithm. Here each iteration requires a modified MVCG and an additional quantum gate. For the $k^{th}$ iteration out of all $n$ iterations, the $\hat{U}$ gate becomes $\hat{U}^x$, where $x={d^{(n-k)}}$, and therefore the modified MVCG applies $(\hat{U}^x)^j$ to the target qudit when the control qudit is $\ket{j}$. As our current approach applies $\hat{U^j}$ directly (i.e. not cascading $\hat{U}$ $j$ times), implementing the MVCG with $(\hat{U}^x)^j$ is no more challenging than the MVCG with $\hat{U^{j}}$. For the additional quantum gate, on the $k^{th}$ iteration the control qudit undergoes an $R_z$-rotation of the angle $\theta=-\sum\limits_{i=1}^{k-1} \frac{\phi_{i}}{d^{k-i}}$ where each $\phi_i$ is the phase determined by the $i^{th}$ iteration prior to the $k^{th}$ iteration. Our successful implementation of the controlled gate in this paper demonstrates all the capability needed for implementing the arbitrary (diagonal) $R_z$-rotation. We note the two operations needed to achieve a standard iterative PEA can also be used to implement a Bayesian phase estimation approach known as rejection filtering phase estimation (RFPE)\cite{Paesani2017}. RFPE is robust to noise and promises a speed up over standard iterative PEAs by gaining information about multiple bits (for us, dits) of the phase at a time. For the standard (non-Bayesian) PEA, qudits provide a $\log_2(d)$ reduction in the number of iterations needed to estimate a given phase with success rate identical to the qubit case. As with qubit systems, an arbitrarily-high success rate can be achieved via multiple trials for some (or all) iterations. As our photonic system provides photon statistics easily, a low-error iterative PEA or RFPE is a natural next step. Using an iterative PEA avoids cumbersome multi-photon gates for the control qudit; however, we note that multiple target qudits may be needed to accommodate a unitary $\hat{U}$ of a high dimension $M$. To be precise, the number of target qudits $m$ must be $\geq \log_d(M)$, thus $\hat{U}$ becomes a multi-photon gate when $\log_d(M)>1$. Our ability to implement a high-dimensional $\hat{U}$ scales polynomially with the qudit dimension $d$ and exponentially with the number of target qudits $m$.

In conclusion, this work has successfully demonstrated the first implementation of the PEA on a qudit-based photonic platform. This experiment utilized the high dimensionality of the time and frequency DoFs on a single photon to realize the 2-qudit MVCG gate, circumventing the inherently probabilistic photon-photon interactions. Although limited to a proof-of-principle model with arbitrary-phase diagonal unitaries, this work is a first physical demonstration of a qudit-based PEA. Future improvements to our PEA include higher-dimensional qudits ($d>3$), arbitrary (non-diagonal) unitaries, and statistical estimation of the phase via larges ensemble measurements.

\section*{Acknowledgements}
This work was supported in part by the National Science Foundation under award number 1839191-ECCS. The authors thank Prof. Minghao Qi for helpful discussions.

\end{document}

%% file: PhaseFittingPlot.tex
\usepackage{xfrac}
\usepackage{pgfplots}

\definecolor{cf0}{rgb}{0,0.4470,0.7410}
\definecolor{cf1}{rgb}{0.8500,0.3250,0.0980}
\definecolor{cf2}{rgb}{0.9290,0.6940,0.1250}

\newcommand{\PhaseFitting}{
	\begin{tikzpicture}[x=1cm,y=6cm, baseline=(current bounding box.center), thick,scale=1, every node/.style={scale=1.3}]
		\tikzstyle{every node}=[font=\scriptsize]
		
		\begin{axis}[
			name = MyAxis,
    		ymin=0, ymax=1.05,
    		xmin=0, xmax=6.4,
    		axis x line=center, 
    		axis y line=center, 
    		xtick={
        		0,    1.0472,    2.0944,    3.1416,    4.1888,    5.2360,    6.2832
    		},
    		xticklabels={
        		$0$, $\sfrac{\pi}{3}$, $\sfrac{2\pi}{3}$, $\pi$,
        		$\sfrac{4\pi}{3}$, $\sfrac{5\pi}{3}$, $2\pi$
    },
        	ytick={
        		0,    0.1111,    0.2222,    0.3333,    0.4444,    0.5556,    0.6667,
        		0.7778,    0.8889,    1.0000
    		},
    		yticklabels={
        		$0$,	$\sfrac{1}{9}$,	$\sfrac{2}{9}$,	$\sfrac{1}{3}$,	$\sfrac{4}{9}$,
        			$\sfrac{5}{9}$,	$\sfrac{2}{3}$,	$\sfrac{7}{9}$,	$\sfrac{8}{9}$, $1$
    },
    		grid = major,
    		xlabel = {\large Phase},
    		ylabel = {\large Probability},
    		xlabel style = {at = {(ticklabel cs:.5)}, anchor=south, yshift=-1.75ex, },
    		ylabel style = {at = {(ticklabel cs:.5)}, anchor=center, xshift=-1.25ex, rotate=90},
    		ticklabel style = {font=\normalsize},
    		legend columns=-1,
    		legend style={at={(0.5,1.1)},anchor=north, draw=none}
]
		\addplot [
			color=cf0,
			thick,
			samples=201,
			domain = 0:6.28
			]
			{(1/9)*(3 + 4*cos(deg(x)) + 2*cos(2*deg(x))) };
			\addlegendentry{$\ket{0}_f$}
		\addplot [
			color=cf1,
			thick,
			samples=201,
			domain = 0:6.28
			]
			{(1/9)*(3 - cos(2*deg(x)) - sqrt(3)*sin(2*deg(x)) - 2*cos(deg(x)) +2*sqrt(3)*sin(deg(x))) };
			\addlegendentry{$\ket{1}_f$}
		\addplot [
			color=cf2,
			thick,
			samples=201,
			domain = 0:6.28
			]
			{(1/9)*(3 - cos(2*deg(x)) + sqrt(3)*sin(2*deg(x)) - 2*cos(deg(x)) -2*sqrt(3)*sin(deg(x))) };
			\addlegendentry{$\ket{2}_f$}
			
		\addplot[mark=none, dash pattern={on 1pt off 1pt}, line width=.4mm, color=green!65!black,] coordinates {(6.2832,1.1) (6.2832,0)};
		\addlegendentry{True $\phi$}
		\addplot[mark=none, dashed, line width=.3mm, color=black!90] coordinates {(5.8386,1.1) (5.8386,0)};
		\addlegendentry{Est $\tilde{\phi}$}
		
		\addplot[mark=halfsquare*, scale=1pt, fill=cf0, mark color=cf0] coordinates {(5.8386, .878)};
		\addplot[mark=halfsquare*, scale=1pt, fill=cf1, mark color=cf1] coordinates {(5.8386, .032)};
		\addplot[mark=halfsquare*, scale=2pt, fill=cf2, mark color=cf2] coordinates {(5.8386, .090)};

		\addplot[mark=halfsquare*, scale=1pt, fill=cf0, mark color=cf0] coordinates {(1.1831, .316)};
		\addplot[mark=halfsquare*, scale=1pt, fill=cf1, mark color=cf1] coordinates {(1.1831, .530)};
		\addplot[mark=halfsquare*, scale=2pt, fill=cf2, mark color=cf2] coordinates {(1.1831, .154)};
		\addplot[mark=none, dash pattern= on 4pt off 3pt, line width=.3mm, color=black!90] coordinates {(1.1831,1.1) (1.1831,0)};
		\addplot[mark=none, dash pattern={on 1pt off 1pt}, line width=.4mm, color=green!65!black,] coordinates {(1.1030,1.1) (1.1030,0)};

		\addplot[mark=halfsquare*, scale=1pt, fill=cf0, mark color=cf0] coordinates {(3.283, .143)};
		\addplot[mark=halfsquare*, scale=1pt, fill=cf1, mark color=cf1] coordinates {(3.283, .318)};
		\addplot[mark=halfsquare*, scale=2pt, fill=cf2, mark color=cf2] coordinates {(3.283, .539)};
		\addplot[mark=none, dash pattern= on 4pt off 8pt, line width=.3mm, color=black!90] coordinates {(3.283,1.1) (3.283,0)};
		\addplot[mark=none, dashed, dash pattern= {on 1pt off 1pt on 1pt off 1pt on 1pt off 7}, dash phase=6.5pt, line width=.4mm, color=green!65!black,] coordinates {(3.284,1.1) (3.284,0)};

		\end{axis}
		
		\end{tikzpicture}

}